\documentclass[sigconf,authorversion,nonacm]{acmart}

\usepackage{balance} 
\usepackage{enumitem}
\usepackage{hyperref}
\usepackage{footnote}
\usepackage{amsmath}
\usepackage{booktabs}
\usepackage{multirow}
\usepackage[font=small,labelfont=bf]{caption}

\title[Stratsphere]{LLM-augmented empirical game theoretic simulation for social-ecological systems}

\author{Jennifer Shi}
\affiliation{
  \institution{Western University}
   \city{London, ON}
  \country{Canada}}
\email{jshi372@uwo.ca}

\author{Christopher K. Frantz}
\affiliation{
  \institution{Norwegian University of Science and Technology}
  \country{Norway}}
\email{christopher.frantz@ntnu.no}

\author{Christian Kimmich}
\affiliation{
  \institution{Institute for Advanced Studies (IHS)}
  \city{Vienna}
  \country{Austria}}
\email{kimmich@ihs.ac.at}

\author{Saba Siddiki}
\affiliation{
  \institution{Syracuse University}
  \city{Syracuse, NY}
  \country{United States}}
\email{ssiddiki@syr.edu}

\author{Atrisha Sarkar}
\affiliation{
  \institution{Western University}
  \city{London, ON}
  \country{Canada}}
\email{atrisha.sarkar@uwo.ca}

\begin{abstract}
Designing institutions for social-ecological systems requires models that capture heterogeneity, uncertainty, and strategic interaction. Multiple modeling approaches have emerged to meet this challenge, including empirical game-theoretic analysis (EGTA), which merges ABM's scale and diversity with game-theoretic models' formal equilibrium analysis. The newly popular class of LLM-driven simulations provides yet another approach, and it is not clear how these approaches can be integrated with one another, nor whether the resulting simulations produce a plausible range of behaviours for real-world social-ecological governance. To address this gap, we compare four LLM-augmented frameworks—procedural ABMs, generative ABMs, LLM-EGTA, and expert guided LLM-EGTA, and evaluate them on a real-world case study of irrigation and fishing in the Amu Darya basin under centralized and decentralized governance. Our results show: first, procedural ABMs, generative ABMs, and LLM-augmented EGTA models produce strikingly different patterns of collective behaviour, highlighting the value of methodological diversity. Second, inducing behaviour through system prompts in LLMs is less effective than shaping behaviour through parameterized payoffs in an expert-guided EGTA-based model.
\end{abstract}

\keywords{Agent-based modelling, Generative agents, Game-theoretic simulation, institutional analysis}

\newcommand{\BibTeX}{\rm B\kern-.05em{\sc i\kern-.025em b}\kern-.08em\TeX}

\begin{document}


\pagestyle{fancy}
\fancyhead{}


\maketitle 


\section{Introduction}
Understanding the governance of complex social-ecological systems is a pressing challenge of our times \cite{gotts2019agent, ostrom2009general}. Policies dealing with water scarcity, energy provision, financial stability, and economic inequality depend on how people behave in settings where individual actions relate to collective outcomes. These systems are often marked by uncertainty, heterogeneity, and strategic interaction, making them difficult to study with traditional structural methods alone. Computational models have therefore become essential for exploring how local behaviour translates into global outcomes, and for testing institutional arrangements before they are implemented in the real world. There are four broad categories of computational methods for this purpose: agent-based modeling (ABM) \cite{gotts2019agent}, game-theoretic analysis \cite{janssen2010lab}, hybrid Empirical Game-Theoretic Analysis (EGTA) \cite{wellman2025empirical} models that combine the two, and more recently, large language model (LLM)-based simulations \cite{anthis2025llm}. In this paper, we evaluate a new approach that aims to synthesise these traditions: LLM augmented EGTA. By embedding large language models within the EGTA pipeline, this approach takes advantage of the ability of LLMs’ to generalise in behavioural contexts, generate plausible strategies, and reason about incentives expressed in natural language; and at the same time, retain the formal precision of game-theoretic analysis. Through comparative experiments in a real-world social-ecological case study, we compare and evaluate LLM augmented EGTA with procedural ABM, generative ABM models in terms of behavioural outcomes, sustainability, and robustness. \par
Agent-based models (ABMs) have been a mainstay in computational approaches to studying collective behaviour for more than a half century \cite{schelling1969models}, with a broad adoption in social-ecological systems (SES) modelling \cite{an2021challenges, gotts2019agent}. ABMs offer a flexible means to represent heterogeneous agents, local interactions, and emergent system-level outcomes, making them especially well-suited for exploring complex adaptive systems. In recent years, the general-purpose capabilities of large language models (LLMs) have given ABMs renewed momentum \cite{anthis2025llm}. The integration of natural language reasoning with agent-based simulation, often referred to as \textit{generative agents}, has enabled the development of models that can generalise, communicate, and interpret context in a manner that procedural ABMs cannot. Such models have already been applied in various domains, including replications of classic social psychology experiments \cite{aher2022using}, analyses of financial markets \cite{gao2024simulating}, and studies of electoral behaviour \cite{argyle2023out}. However, Generative ABMs still largely emphasise descriptive realism and behavioural plausibility rather than strategic reasoning and equilibrium stability. \par
Game-theoretic simulations occupy a different point in this methodological landscape. They replace the heterogeneity and scale of ABMs with formal reasoning about equilibria. However, traditional analytical game-theoretic models often require strong simplifying assumptions about the number of players, payoff structures, or available strategies. Empirical Game-Theoretic Analysis (EGTA) \cite{wellman2025empirical} bridges this gap by combining simulation-based data generation through ABM with game-theoretic reasoning. In EGTA, simulation is used to estimate payoffs empirically and then formal analysis is applied to identify equilibrium strategies. This approach allows for the study of strategic interaction in systems that would otherwise be analytically intractable while retaining the formal rigour.

Despite the strengths of these existing approaches—ABM, generative ABM, game-theoretic simulation, and EGTA—each occupies a distinct methodological niche. ABMs excel at exploring emergent behaviour but struggle with analytical generalisation. Game-theoretic models offer formal guarantees, but at the cost of ecological realism. EGTA balances the two, yet still depends on designer-specified behavioural models that can be difficult to generalise beyond specific settings.

In summary, we make three main contributions, each addressing a distinct research question:\par
\noindent \textit{RQ.1:} \textit{How can the generalisation and reasoning capabilities of LLMs be integrated with EGTA-based simulation?} To that end, we propose (Sec~\ref{sec:frameworks}) two approaches: (i) a fully automated framework that augments each component of the EGTA pipeline with LLMs, and (ii) an expert-guided framework, where expert oversight refines and corrects LLM-generated game models within the pipeline. \par

\noindent \textit{RQ.2:} \textit{How do these simulation approaches perform in practice?} We answer this by applying them to a real-world case study of irrigation and fishing in the Amu Darya basin under centralised and decentralised governance (Sec~\ref{sec:case_study}). We compare results in terms of community wealth, economic activity distribution and resource sustainability, and benchmark against procedural ABMs and LLM-driven generative agents.\par

\noindent \textit{RQ.3:} \textit{How sensitive are these approaches to the choice of language model and the induction of agent behaviour?} To assess this, we replicate the case study using three different LLMs, each induced with three behavioural profiles (rational, altruistic, and balanced), and compare their results with those generated through incentive-based behavioural induction using Pigouvian taxes (Sec~\ref{sec:results}).
 \begin{figure}[t]
    \centering
    \includegraphics[width=0.35\textwidth]{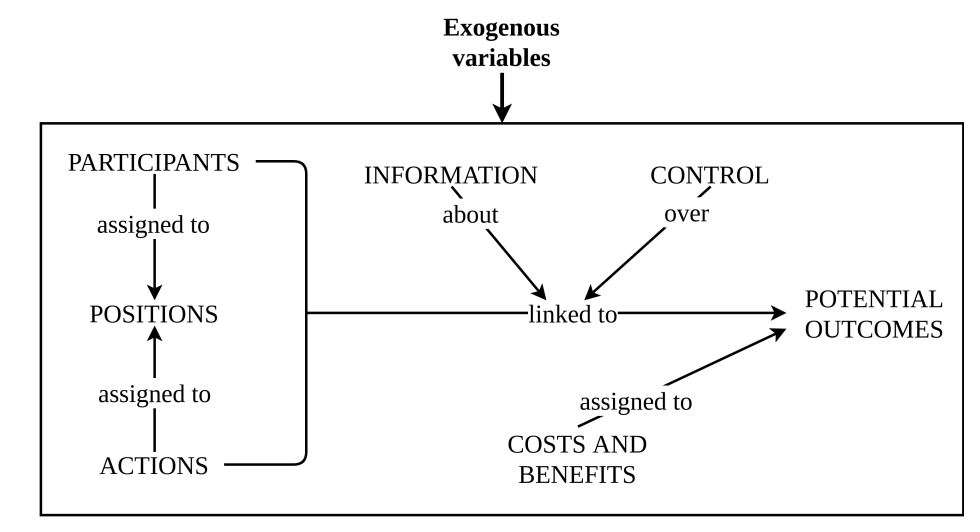}
    \caption{Internal structure of an action situation reproduced from Ostrom 2005~\cite{ostrom2005understanding}}
    \label{fig:as_internal}
\end{figure}
\section{Background}
\subsubsection*{Overview, Design Concepts, Details, and Decision Making (ODD+D)} We draw from the best practices in the ABM community towards standardization and interoperability of computational models. To that end, we use a standardized interface, `ODD+D' (Overview, Design Concepts and Details + Decision Making) \cite{muller2013describing}, as the common input to all the computational models. ODD+D is a textual language for describing the elements of a social-ecological system, including details of the environment and the human behavioural decision-making model that an ABM simulates. The language provides a standardized description of the system under study and helps reproducibility and a clear understanding of the assumptions built into the model. Key elements of ODD+D include a description of the purpose of the model, the entities and state variables, the individual decision-making, learning, and prediction models, and the parameters that capture heterogeneity and stochasticity. For a detailed account of the protocol, we refer to Müller et al.~\cite{muller2013describing}; the specific ODD+D used in our case studies is included in the supplementary material. While the original motivation for ODD+D was the standardization of ABM descriptions, its textual format also provides a convenient structure for defining the interface for an LLM-augmented simulation system. Moreover, this standardized interface can remain invariant across different design approaches. For example, the structure can be used as a context for generating code in traditional procedural ABMs, for generative ABMs, or as a preliminary step for extracting information about the environment in an EGTA-based approach.\par
\small
\begin{figure*}[!ht]

    \centering
    \includegraphics[width=0.85\textwidth]{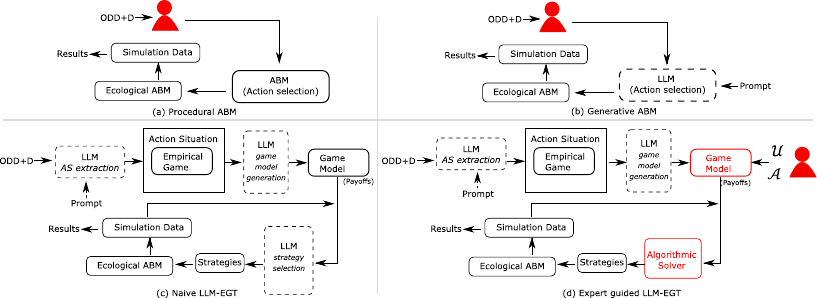}
    \caption{Pipeline for four different simulation methodologies. Panels (c) and (d) show two approaches to LLM augmentation within the EGTA pipeline: a naive approach that relies solely on LLMs at each step, and an expert-guided approach in which a designer refines the game model. The ecological ABM, which simulates the environmental and ecological variables, is integrated into all four approaches.}
    \label{fig:model_schemas}
\end{figure*}
\normalsize
\subsubsection*{Action Situation}
 Ostrom's Institutional Analysis and Development framework \cite{ostrom2005understanding} builds on the conception of situational arrangement of actors in a given social or resource dilemma and captures their social interactions, specifically their action choices and associated outcomes in the form of so-called `Action Situations' as the basic unit to analyze institutional arrangements within a complex social system. Action Situation (AS) first defines an abstraction of the interactions; for example, a community of farmers facing a collective action problem of a common-pool resource exploitation dilemma. ASs also provide a way to structure the game that represents the interaction between the participants \cite{montes_computational_2022} and consequently a way to analyse strategies under different behaviour models of decision making, explore changes in the rule and game structure, and provide a framework to a policy analyst for answering `what if' questions for proposed policies. EGTA and various other tools can be employed as a computational approach to explore such questions. Typically, the identification of potential AS(s) in a social-ecological system is performed by the policy analyst; often through on-field interviews and additional forms of expert knowledge based on recurrent archetypical patterns of social-ecological problems \cite{oberlack2019archetype} and archetypical games \cite{bruns_archetypal_2021}. Although expert knowledge is indispensable in this process, LLMs can potentially also play a valuable role. If the elements of the system are described completely in an ODD+D specification, LLM should be able to extract, in theory, the underlying AS(s) in the system, which can then be further refined by the policy analyst. 
 \subsubsection*{Empirical Game-theoretic analysis} Empirical Game-Theoretic Analysis (EGTA) \cite{wellman2006methods} integrates the heterogeneity and scalability of Agent-Based Model (ABM) simulations with the formal structure of game-theoretic modelling, thus supporting formal equilibrium analysis. The framework is defined by a few essential components \cite{wellman2025empirical}. First, since a real-world system is often too complex to be represented with a single game, EGTA factors this system into a manageable set of \textit{empirical games}. For example, a complex N-player sequential auction might be replaced by a set of 2-player games where each player $i$ controls one agent and views all other players ($-i$) as an aggregate entity rather than as individuals. The second component is the game model itself, which involves estimating the payoffs of these empirical games. These payoffs are typically estimated through a variety of methods, including machine learning, heuristics, or by running an ABM simulation under various combinations of agent strategies. Finally, a solution concept is used for solving for the strategies of the game. Any standard concept can be employed; we use Nash equilibrium. One novelty of our approach is that it integrates AS and EGTA by using ASs to identify empirical games within the EGTA pipeline.

\section{Simulation methodologies}
\label{sec:frameworks}
\subsection{LLM-augmented EGTA}
Fig.~\ref{fig:model_schemas}(c) and (d) illustrates two distinct approaches to incorporate LLM-based components within an empirical game theoretic simulation (EGTA) pipeline. Both aim to extend traditional EGTA by automating or augmenting key steps, such as game construction, strategy generation, and payoff estimation, through the reasoning and generalization capabilities of LLMs. The difference lies in the degree of autonomy granted to the LLM and the extent of human oversight in the simulation process. 
\subsubsection{Naive LLM-EGTA}
In this model, every step within the EGTA framework \cite{wellman2025empirical} is augmented with an LLM. In Step~1, we merge both the extraction of action situations (AS) from the ODD+D description and the generation of the corresponding game models (Fig.~\ref{fig:model_schemas}c) into a single step. The LLM is tasked with inferring the ASs and identifying the structure of the embedded game model within each extracted AS. To assist the LLM in selecting an appropriate game model, we include a set of canonical games \cite{bekius2020selecting} in the prompt. Examples include the Public Goods Game and the Common Pool Resource Game, among others (all prompts are provided in the supplementary material). In Step~2, the LLM receives additional information about the state of the ecological variables (see Sec.~\ref{sec:ecological_abm}) and is prompted to select a strategy for each game. Once the strategies of all actors are generated, the ecological ABM simulates the dynamics of the ecological variables to produce the corresponding simulation data, which is then used to update the system state and advance to the next time step.

\subsubsection{Expert-guided LLM-EGTA}
The pipeline for the expert-guided approach to LLM augmentation in EGTA differs from the naive version in two main ways, as illustrated in Fig.~\ref{fig:model_schemas}d. First, the expert retains greater control over the game model. The game models generated by the LLM may not always be realistic for the problem domain; in such cases, the designer can refine them by incorporating domain-specific knowledge. For example, if the empirical game in an AS represents a common-pool resource problem, the LLM might propose agent actions as a binary choice between “high” and “low” extraction levels, whereas the designer may prefer a continuous formulation with a closed-form payoff function defined over a continuous strategy profile. This design allows such refinements to be implemented directly within the simulation loop, rather than being constrained by the LLM-generated structure. Second, because the expert can introduce more complex game models in this approach, we employ reliable algorithmic solvers such as Gambit \cite{savani2023gambit} instead of relying on the LLM to compute equilibrium strategies. This modelling framework is particularly well suited for scenarios too complex to depend solely on the LLM’s capacity to solve sophisticated game structures.
\vspace{-0.5em}
\subsection{ABM-based approaches}
In addition to the LLM-augmented EGTA models, we implement two ABM-based methodologies that serve as comparative baselines: a \textit{procedural ABM} and a \textit{generative ABM} (Fig.~\ref{fig:model_schemas}a,b). Both share a common ecological simulation environment but differ in how agents select actions. In the \textit{procedural ABM}, agent actions are determined by predefined behavioural functions coded in a procedural language. These functions map environmental and social variables to specific actions based on rules specified by the model designer. In the \textit{generative ABM}, by contrast, the action selection process is delegated to an LLM. We adopt a role-based prompting approach \cite{gurcan2024llm}, in which each agent is assigned a role (e.g., farmer) and a behavioural profile, along with contextual information about the state of the environment. The LLM is then queried to determine the agent’s next action. Since the details of agent behaviour and environmental dynamics are case-specific, the full behavioural specifications are discussed in Sec.~\ref{sec:behaviour_models}.

\section{Case Study}\label{sec:case_study}
We now present the real-world case study we use for evaluating and contrasting the differences between the two LLM-augmented EGTA approaches, along with the generative agent approach, and procedural ABMs. The case focusses on irrigation allocation in the Amu Darya basin, a well-studied social-ecological system characterised by interdependent water use and strategic decision-making among farmers \cite{Schlter2005, Ziganshina2022}. 
 \begin{figure}[t]
    \centering
    \includegraphics[width=0.4\textwidth]{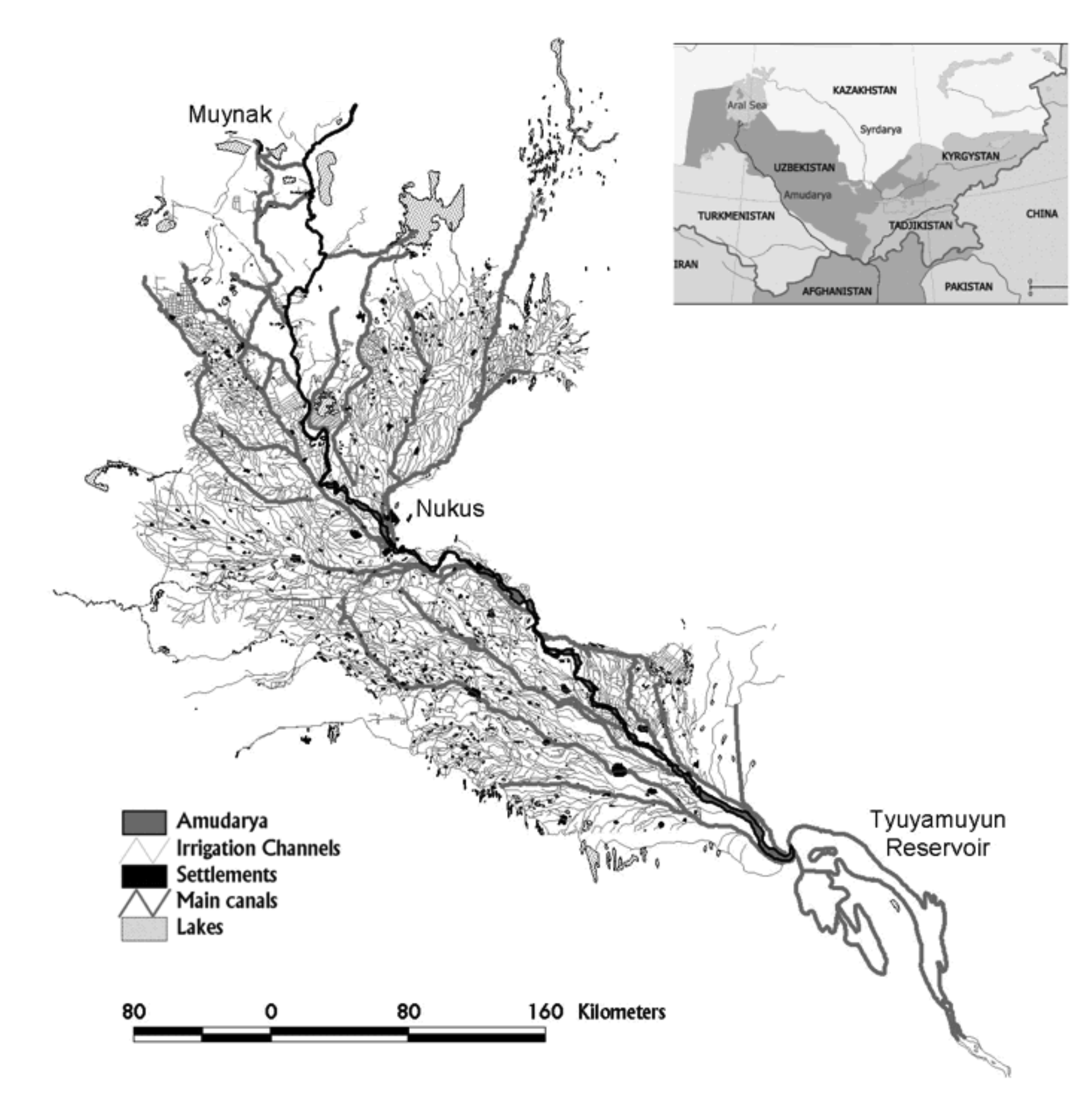}
    \caption{Map of Amu Darya Basin reproduced from \cite{schluter2007mechanisms}.}
    \label{fig:amu_darya_map}
\end{figure}
The social-ecological environment captures the interaction between individual farmers, a river basin, downstream lake ecosystem, and two different institutional governance approaches, namely centralised and decentralised. We describe the key elements of the ecosystem below, and refer to Schlüter and Pahl-Wostl \cite{schluter2007mechanisms} for more details. 

\noindent \textit{Farmers:} Farmers are spatially located in fixed positions along the river basin (Fig. \ref{fig:amu_darya_map}), and they decide annually on the number of fields to irrigate. Higher withdrawal decisions negatively affect downstream farmers due to reduction in the water flow, and also negatively affects the available fish population year over year due to reduction of water in the lake. If there is not enough water flow to sustain the number of fields the farmer has decided to irrigate, the reduced water availability negative affects the crops due to plant stress, which accumulate over time. The location of a farmer determines its access to water flowing from the reservoir (upstream farmers have easier access) and access to the downstream lake for fishing (downstream farmers have easier access). The yields from the irrigated fields and fishing contribute to the annual budget of the farmer, part of which they reinvest for planting crops in the subsequent seasons. Each farmer's decision of the number of fields they choose follows the behaviour model that each simulation framework uses (Sec~\ref{sec:behaviour_models}); however, examples of variables that influence their decision are their expectations of water availability based on previous flows, their memory of previous crop yields, and their annual budget. Their access to fishing is determined by their proximity to the lake and upstream farmers has access to only the stock remained after downstream farmers are done fishing upto their target catch. Fishing time has no bearing on time available for farming (imagine one member of a household engages in the farm, while the other in fishing). 

\noindent \textit{National authority:} A national regulatory authority has a role in the centralized regime; it
predicts water inflow, determines the number of fields that can be
supported based on budget constraints, and allocates fields equally
to all farmers. Its budget is determined by the cumulative agricultural returns, minus irrigation and consumption costs. 

\noindent  \textit{Fishes:} The fish population inhabits a terminal lake at the end of the river. There are thirteen classes of fish across different ages of maturity:
one for larvae (age 0), four juvenile classes (ages 1–4), and eight
adult classes (ages 5–12). Larvae enter the population either via
reproduction by mature adults or through migration from upstream,
which depends on whether inflow into the lake during a specific month of the year (May) exceeds a minimum ecological threshold. Mortality of fish in the juvenile stages is Malthusian because of limited resources. Farmers harvest fish from adult age classes in order
of proximity to the lake.

\noindent \textit{River:} Water in the river enters the system from upstream
and each farmer extracts water according to their irrigation plan proportional to the number of fields they decide to irrigate for the year, with any remaining volume passed downstream. The terminal lake
receives the outflow, which in turn affects fish reproduction and
abundance. \par

\subsection{Simulation of ecological variables}\label{sec:ecological_abm}
Operationalisation of the simulation pipelines, as shown in Figure \ref{fig:model_schemas}, requires modelling the dynamics of the above ecological variables. We use a procedural ABM for ecological variables that simulates the following variables: (i) crop yields of each farmer based on the number of irrigated fields, (ii) monthly river flow, (iii) size and distribution of the fish population, and (iv) accumulated monetary returns from farming and fishing for each farmer as well as global returns that represent the budget of the national authority. The equations governing these models are included in the supplementary material.

Each simulation year follows a sequence of steps that reflect seasonal irrigation activities, ecological dynamics, and the resulting economic effects. We implement the simulation loop over a 100-year horizon. For farmer behaviour, we simulate two governance modes, following Schluter \cite{schluter2007mechanisms} -- decentralized and centralized. Farmer actions that are guided by the four approaches presented in Sec. \ref{sec:frameworks} are all decentralized approaches since farmers make their own decisions. Centralized approaches are implemented for baseline comparison using a procedural ABM in which the national authority is responsible for water allocation. 
\section{Behaviour models}\label{sec:behaviour_models}
\subsection{Decentralized models} In the decentralized governance model, there is no centralized authority; each farmer independently determines irrigation strategy using one of the four models in Sec. \ref{sec:frameworks}.
\subsubsection{Procedural ABM:} We use an identical setup as Schluter \cite{schluter2007mechanisms} for the procedural ABM. Farmers use a rule-based approach informed by past yields and a prediction of estimated water availability. Farmer are modelled with bounded memory, and use the previous year's inflow to predict the upcoming year's inflow. If the previous year’s yield is below the subsistence level, the farmer increases the number of irrigated fields by one (if they have sufficient reserves), or otherwise calculates the number of fields by dividing financial reserves by the irrigation cost per field. If the previous year’s yield is above subsistence level, the farmer calculates the number of fields by dividing the estimated water availability by the per-field water requirement. Exogenous parameters, such as per-field water needs and irrigation costs, follow the specification in \cite{schluter2007mechanisms}.
\subsubsection{Generative ABM:} The structure of generative ABM is similar to procedural ABM, but the rule-based decision-making is replaced by an LLM agent. The LLM is given a role of a farmer along with a behavioural induction (altruism, balanced, rational) using the system prompt and provided with information on predicted water availability, past yield, minimum subsistence income, per-field water requirements, and irrigation costs, and is tasked with deciding the optimal number of fields to irrigate for the season.
\subsubsection{Naive LLM-EGTA simulation:} \label{sec:llm_egta_descr} In the naive LLM-EGTA model, in the first step, an LLM is given the structure of an AS and prompted to construct them from the ODD+D description. 
Although the models can generate any number of ASs, we select two based on the correct interpretation of the environment: the first AS consists of a 2-player cooperation game between neighbouring farmers that models the water withdrawal decisions of the farmers along the river; the second AS represents an $N$-player common pool resource (CPR) game to model the amount of fish extraction for each farmer. The LLMs used for this step (DeepSeek-V3, 671B param) was able to successfully extract both sets of AS in repeated simulation runs. In the next step, the LLM is prompted with a role as a farmer and each AS as additional context to select the actions in the two sets of games that each farmer plays. Once each farmer's actions are generated, the rest of the ecological variables are modelled using the ecological ABM described earlier. \par
\subsubsection{Exp LLM-EGTA simulation:} \label{sec:dil_llm_egta_descr} The \emph{expert-guided} approach follows the naive LLM-EGTA simulation with the designer refining the elements of the generated game model. Specifically, in the naive LLM-EGTA model, the game model constructed by all the LLMs are abstract actions, such as high/low numbers of irrigation fields (corresponding to high and low water withdrawal) instead of more granular action space of integers representing the number of fields and fished for each of the games. Therefore, we change the game models to a more granular level of actions to be solved by an algorithmic game solver. For the fishing AS, we use a standard common pool resource game with a range of integer extraction levels corresponding to the number of fishes to extract, and the details of the farming AS game is described as follows:\\
Let $s_u, s_d \in [0,\text{max fields}]$ denote the upstream and downstream farmers’ strategy choices, respectively. The upstream farmer has priority access to irrigation water, with the downstream farmer receiving water flow after the upstream farmer’s withdrawal. The number of fields actually irrigated for each farmer is therefore determined by affordability constraints (annual budget and per-field costs), physical water availability, and river flow. Formally, the irrigated fields are defined as
\[
f_u = \min\!\left(s_u,\; \frac{B_u}{c},\; \frac{T}{w}\right), \qquad
f_d = \min\!\left(s_d,\; \frac{B_d}{c},\; \max\!\left(\frac{T}{w} - f_u,\,0\right)\right),
\]
where $B_u, B_d$ denote the upstream and downstream farmers' budgets from the accumulated returns of farming and fishing from previous years, $c$ is the irrigation cost per field, $T$ is the total water available in the river, and $w$ is the annual water requirement per field. The ecological stress on crops due to the same amount of water distributed across a large number of fields is modelled by applying a yield reduction once the total number of irrigated fields exceeds a stress threshold $S$. The per-field yield is thus defined as
\[
y = 
\begin{cases}
y_0, & f_u + f_d \leq S,\\
y_s, & f_u + f_d > S,
\end{cases}
\]
where $y_0$ is the baseline yield and $y_s < y_0$ is the stressed yield. Each farmer’s payoff is given by the resulting agricultural yield and fish income, minus the irrigation and consumption costs:
\begin{align}
\begin{split}
\pi_u(s_u, s_d) &= f_u \, y + F_u - \big(f_u c + \kappa\big)  - \tau \cdot (s_{u}+s_{d}) \cdot s_{u} \\
\pi_d(s_u, s_d) &= f_d \, y + F_d - \big(f_d c + \kappa\big) - \tau \cdot (s_{u}+s_{d}) \cdot s_{d}
\label{eqn:dil_llm_EGTA_eqn}
\end{split}
\end{align}
where $F_u, F_d$ are fishing incomes from the second AS and $\kappa$ is the fixed consumption cost, and $\tau$ is the tax that internalizes the social cost of extraction into the payoffs. $\tau > 0$ represent Pigouvian taxes, whereas $\tau = 0$ represent a purely self-interested player. The Nash equilibrium $(s_u^*, s_d^*)$ satisfies 
$\pi_u(s_u^*, s_d^*) \ge \pi_u(s_u, s_d^*)$ and 
$\pi_d(s_u^*, s_d^*) \ge \pi_d(s_u^*, s_d)$ 
for all $s_u, s_d \in [0, \text{max fields}]$.
 We use the above payoff functions to construct the normal-form game played between pairs of consecutive farmers and solve for the Nash equilibrium strategies using the Gambit solver \cite{savani2023gambit} implementation of the Lemke–Howson algorithm \cite{lemke1964equilibrium}. Since each farmer, other than those at the start and end, participates in two games, we randomly select one of the strategies generated from the two. These strategies are then used as the number of fields irrigated by each farmer in the EGTA pipeline, which subsequently feeds into the ecological ABM simulation for the remaining ecological variables.
\subsection{Centralized model}
In the centralized version, farmers play no role, and the national authority is responsible for water allocation. The authority predicts annual water availability using memory of past inflows using a 20 year moving average, and then determines the number of irrigable fields based on budget constraints, and allocates these fields equally among farmers. The simulation is primarily used for comparative analysis of the decentralized models.

\section{Results}\label{sec:results}
 \begin{figure*}[t]
    \centering
    \includegraphics[width=0.68\textwidth]{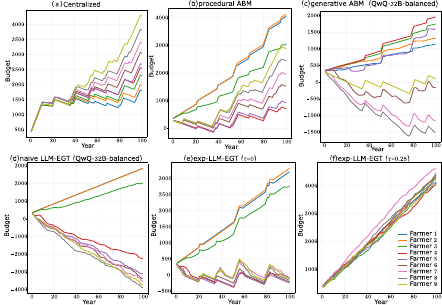}
    \caption{Total budget (including farming and fishing) of each farming household  from upstream (1) to downstream (9) over a period of 100 years under different simulation models and frameworks. $x$ axis: time in years, $y$ axis: household wealth.}
    \label{fig:farmer_budgets}
\end{figure*}
We now present how the results of the simulation differ with respect to the overall budget of each farming household, the activity distribution of individual farmers, and robustness of the models to underlying LLMs and behavioural inductions.  Each simulation is run using a calibrated real water inflow observed at the start of the season (July). All exogenous parameters are kept identical across behaviour models and are calibrated to the ones used in Schluter \cite{schluter2007mechanisms}. We run the following sets of models for our comparative evaluation: (i) Centralized allocation model, (ii) a procedural ABM, (iii) generative ABM, (iv) Naive LLM-EGTA, (v) Expert-guided LLM-EGTA with $\tau=0$, i.e., no Pigouvian taxes, and (vi) Expert-guided LLM-EGTA with $\tau=0.25$, i.e., with Pigouvian taxes.

 \begin{figure*}[!th]
    \centering
    \includegraphics[width=0.85\textwidth]{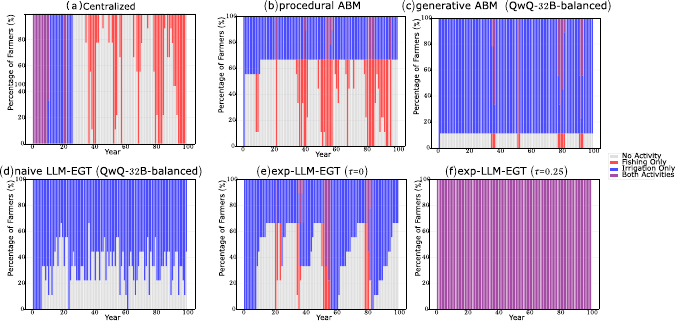}
    \caption{Distribution of activities across farming households in the population over a period of 100 years under different simulation frameworks. Each horizontal bar is a stacked-bar chart showing the percentage of households ($y$ axis) that engage in each activity.}
    \label{fig:activity_summary}
\end{figure*}

\vspace{-0.5em}
\subsubsection*{Farmer budget:} Fig. \ref{fig:farmer_budgets} shows the cumulative year-over-year budget (y-axis), calculated as the sum of farming and fishing income, for each farmer household over time (x-axis). The first observation is that the total wealth varies significantly between the different models. As a reference, panel (e) shows household budgets under the Exp-LLM-EGTA simulation with $\tau=0$ (without the Pigouvian tax). Here, we clearly see a tragedy-of-the-commons effect: other than three upstream farmers, all households fall into debt (negative budgets) by the 10-year mark, while upstream farmers continue to accumulate wealth. In this model, farmers irrigate the number of fields predicted by the Nash equilibrium of the CPR game, but the inflow can only sustain the three upstream farmers. Once they withdraw, there is not enough water for downstream irrigation or to maintain the adult fish population in the lake.

We also observe that under the naive LLM-EGTA model, the downstream household debt is much worse than in the Exp-LLM-EGTA (no taxation) model. In naive LLM-EGTA, farmers repeatedly irrigate fields each year incurring a cost, ignoring their shrinking budgets and thereby modelling an almost pathological strategy. This reflects a limitation of purely LLM-based EGTA models: the game generated by the LLM is often a two-action anti-coordination game (e.g., $\bigl( \begin{smallmatrix}6,6 & 5,7\\ 9,3 & 5,2\end{smallmatrix}\bigr)$ corresponding to 'high' and 'low' number of fields ); and even after being prompted to select a granular action in 0 to 10 number of fields, the models can never reason to select 0 under a 'low' strategy. By contrast, in the Exp-LLM-EGTA (no taxation) model, the mathematical formulation of the payoff function by the designer provides a more granular game with actions in the range [0,10] and dynamically changing payoffs based on yearly budgets. As a result, downstream household debt stays close to zero. Interestingly, we see similar dynamics in the generative agent model (panel c), even though the situation is not explicitly presented as a strategic game. However, the chosen actions mirror self-interested strategies instead of ones that account for the social cost. Among the models that sustain the farming community, the Exp-LLM-EGTA with Pigouvian taxation performs best in terms of overall wealth. Centralised allocation and procedural ABM models also support the community, although with higher inter-household inequality. Answer to which household has most wealth varies across models. In models that exhibit the commons-tragedy, upstream households consistently do better, as expected from their preferential water access. By contrast, downstream households fare better in the Exp-LLM-EGTA (Pigouvian taxation) and centralized allocation models. In these cases, egalitarian allocation practices of farms and preferential access to the lake for fishing help downstream households to benefit in comparison to upstream farmers.

These results demonstrate that LLM-augmented EGTA approaches (both naive and expert guided ones) produce dynamics that differ highly from procedural and generative ABMs. At one extreme, naive LLM-EGTA leads to near collapse, while at the other, Exp-LLM-EGTA with Pigouvian taxation sustains the entire community. These distinct dynamics of the LLM-augmented approaches is also reflected when we analyse the distributon of economic activies next.
\vspace{-0.5em}
\subsubsection*{Distribution of activity} Fig. \ref{fig:activity_summary} shows the distribution of farming and fishing activity in the community over time under each simulation model. Each vertical bar represents a stacked bar showing the percentage of households ($y$ axis) in the community that engage in a specific activity in a given year ($x$ axis). Blue represents farming activity, red represents fishing, purple represents both fishing and farming, and grey represents no activity. We see that the patterns of activity are strikingly different across simulation models. Exp-LLM-EGTA (Pigouvian taxes) is the only model that sustains a combined agriculture-and-fishing economy, since in that model, due to the added taxes, the strategies of the farmers stay within the carrying capacity of the river and the lake. Under centralized allocation, which is the second most effective based on overall wealth, we observe distinct phases of activity. Initially, all households start with a combined economy; however, as fish deplete in the river due to higher allocation of irrigation fields, the economy shifts to being farming-based. Following this change, the inflow cannot sustain the determined number of fields, leading slowly to a collapse in which no activity takes place and farmers survive on the accumulated wealth from previous years. Once the flow to the lake recovers, the community can sustain intermittent periods of a fishing-based economy, but the inflow is not enough to sustain a farming-based one. In this model, the collapse of farming can be attributed to the misprediction by the centralized authority of the yearly inflows. In the remaining models, we see a divergence of activities between the upstream and downstream farmers. Whereas the upstream farmers can sustain a farming-based economy in all four, the downstream farmers can only sustain periods of a fishing-based economy (procedural ABM) or no activity at all (LLM-EGTA). Again, we see that LLM-augmented EGTA approaches produce strikingly different patterns, whereas, in this case, generative ABMs show patterns that are somewhat similar to procedural ABMs.

\begin{table}[!ht]
  \centering
  \small
  \setlength{\tabcolsep}{3.5pt}
  \renewcommand{\arraystretch}{0.95}
  
  \begin{tabular}{
    p{1.1cm} c p{1.1cm}|
    r r|
    r p{0.5cm}
  }
    \toprule
    & & &
    \multicolumn{2}{c}{\textbf{Budget (Y 100)}} &
    \multicolumn{2}{c}{\textbf{Activity $\%$}} \\
    \cmidrule(lr){4-5} \cmidrule(lr){6-7}
    \textbf{Simulation} & \textbf{LLM} & \textbf{Induced behaviour} &
    \textbf{Min} & \textbf{Max} &
    \textbf{Both} & \textbf{Irrig.\ only} \\
    \midrule

    \multirow{9}{*}{Gen. agent}
      & \multirow{3}{*}{DeepSeek-R1}
        & altruistic & -1203.78 & 1638.01 &  6.0 & 78.3 \\
      & & balanced   & -2792.10 & 1722.00 &  6.7 & 92.6 \\
      & & rational   & -7133.42 & 2850.00 &  0.0 & 97.4 \\
      \addlinespace[2pt]
      & \multirow{3}{*}{gpt-oss-20b}
        & altruistic & -2249.83 & 1782.00 &  2.4 & 80.8 \\
      & & balanced   & \textbf{-1437.62} & 1572.08 &  9.0 & 89.8 \\
      & & rational   & -5249.93 & 2494.00 &  0.1 & 67.4 \\
      \addlinespace[2pt]
      & \multirow{3}{*}{QwQ-32B}
        & altruistic & \textbf{-969.83} & \textbf{2192.84} & 10.1 & 89.8 \\
      & & balanced   & -1504.45 & \textbf{1953.09} &  7.2 & 81.8 \\
      & & rational   & \textbf{-1214.85} & \textbf{3250.00} &  2.6 & 43.4 \\
    \addlinespace[4pt]
    \midrule

    \multirow{9}{*}{LLM-EGTA}
      & \multirow{3}{*}{DeepSeek-R1}
        & altruistic & \textbf{-668.21} & 1836.68 &  7.6 & 74.0 \\
      & & balanced   & -398.96 & 1716.84 &  7.6 & 78.3 \\
      & & rational   & -3874.02 & 2814.00 &  0.0 & 68.1 \\
      \addlinespace[2pt]
      & \multirow{3}{*}{gpt-oss-20b}
        & altruistic & -3121.95 & \textbf{2138.00} &  0.7 & 61.3 \\
      & & balanced   & \textbf{-277.82}  & 1913.79 &  9.3 & 79.1 \\
      & & rational   & \textbf{-894.69}  & \textbf{2871.91} &  3.2 & 45.4 \\
      \addlinespace[2pt]
      & \multirow{3}{*}{QwQ-32B}
        & altruistic & -1156.71 & 1847.03 &  5.7 & 75.6 \\
      & & balanced   & -3874.02 & \textbf{2850.00} &  0.0 & 68.2 \\
      & & rational   & -3874.02 & 2850.00 &  0.0 & 68.2 \\
    \addlinespace[4pt]
    \midrule

    \multirow{3}{*}{Exp LLM-EGTA}
      & \multirow{3}{*}{}
        & $\tau=$1     & \textbf{3950.00} & \textbf{4550.00} & 100.0 &  0.0 \\
      & & $\tau=$0.25  & \textbf{4041.99} & \textbf{4620.00}  & 100.0 &  0.0 \\
      & & $\tau=$0  & \textbf{-246.01} & \textbf{3310.00} &  6.3 & 50.8 \\
    \bottomrule
  \end{tabular}
  \caption{Robustness analysis: min and max budgets (Year 100) and partial activity breakdown. Highest min and max budget for each simulation and behaviour combination is shown in bold.}
  \label{tab:robustness}
\end{table}
\vspace{-0.5em}

\subsubsection*{Robustness to language models and behavioural inductions}
In this section, we analyze the sensitivity of the models that include LLM augmentation (generative-agent and LLM-EGTA approaches) with respect to two factors. First, we evaluate their sensitivity to the underlying language model itself. For this, we use three open-source reasoning models: DeepSeek-R1, Qwen QwQ-32B, and OpenAI GPT-OSS-20B. Second, we assess the sensitivity of the results to behavioural induction. Recent work by Xie et al.~\cite{xie2025using} demonstrates that LLMs can be induced to replicate a range of behaviours in strategic interactions through behavioural profiles specified via system prompts. Following this approach, we induce three types of behavioural profiles—\textit{altruistic}, \textit{balanced}, and \textit{rational}—for each language model. For example, the rational behavioural induction uses the prompt: \emph{You are a purely self-interested player who always seeks to maximize your own gain and ensure that the outcome is as favourable as possible for yourself.}  (See Supplementary Materials for the other behavioural codes.) Table~\ref{tab:robustness} presents the robustness results for both the generative-agent and LLM-EGTA models. 

As expected, behavioural induction works: we observe that self-interest-driven, rational water extraction leads to faster ecosystem collapse and lower aggregate wealth compared to altruistic or balanced behaviour. This effect is consistent across all models, as the minimum household budget for altruistic and balanced inductions is higher than for rational ones. Overall, these findings indicate that LLM-augmented simulation models are indeed sensitive to behavioural parameters, supporting their use as plausible methodologies for generating interpretable macro-level effects and hypotheses. Such parameters can be empirically estimated from behavioural data or intentionally controlled by the model designer. However, the only exception is QwQ-32B, which produces identical strategies for balanced and rational for naive LLM-EGTA model.

When comparing the naive LLM-EGTA with the Exp-LLM-EGTA model, we observe that the former is considerably less responsive to behavioural induction. In the Exp-LLM-EGTA framework, the parameter $\tau$ controls the extent to which the externality of social costs is internalized in the agents’ payoffs. Pigouvian taxation, mediated by $\tau$, shifts the Nash equilibrium from a purely self-interested strategy ($\tau = 0$), analogous to the rational behavioural induction in LLM-based models, toward more socially balanced strategies ($\tau = 0.25$) and ultimately toward altruistic equilibria ($\tau = 1$). Comparing these two forms of behavioural modulation, we find that both the minimum and maximum household wealth are substantially higher in the Exp-LLM-EGTA model with $\tau = 0.25$ and $\tau = 1$ than in the LLM-based models with balanced and altruistic inductions. This suggests that prompt-based behavioural induction alone is insufficient to generate sustainable behavioural strategies. 

Finally, when analyzing the resulting patterns of economic activity distribution, we find that changing the language model or behavioural induction does not enable the generative-ABM or naive LLM-EGTA models to sustain a combined agriculture-and-fishing economy; all remain confined to predominantly farming-based activity. By contrast, the Exp-LLM-EGTA model with Pigouvian taxation sustains a mixed economy for both $\tau = 0.25$ and $\tau = 1$, while removing the tax ($\tau = 0$) collapses it into a monoculture economy similar to the other models.

\vspace{-0.5em}
\section{Related work}
\noindent \textit{EGTA applications:} Empirical Game-Theoretic Analysis (EGTA) methods have been applied to a wide range of domains, from auctions, financial markets, to cybersecurity. In the realm of auctions and financial markets, EGTA has been employed to study bidding strategies in simultaneous second-price sealed-bid (SimSPSB) auctions, where researchers used the methodology to evaluate strategies that account for complex price dependencies \cite{mayer2013accounting} . Similarly, EGTA has been used to model strategic market choice between continuous double auctions (CDA) and frequent call markets (CALL) for fast and slow traders, which revealed that slow traders generally achieve higher welfare in the CALL market and identified a predator-prey dynamic where fast traders pursue slow traders into either market \cite{wah2016strategic}. In cybersecurity, EGTA was utilized to formulate a Distributed Denial-of-Service (DDoS) attack scenario (the MOTAG Game) as a two-player normal-form game, and assess the strategic stability of various Moving Target Defense (MTD) policies \cite{wright2016moving}. Finally, EGTA has also been applied to analyze financial credit networks (FCNs), modeling the conflict between a payer's interest in minimizing costs during strategic payment routing and the overall network's need for liquidity \cite{cheng2016strategic}. However, the main contributions of this paper, vis-à-vis, an LLM-augmented approach to EGTA, application to social-ecological systems, and comparison of how the strategies differ between traditional and generative ABMs, are all novel related to the existing work, to our knowledge. Outside of EGTA, one recent work has explored categorical game theory as an approach to SES \cite{frey2023composing}.

\noindent \textit{Generative ABM:} Large Language Model (LLM)–based social simulation is rapidly emerging as a novel methodology for agent-based modeling in the social sciences; see \cite{gao2024large, mou2024individual} for a survey. However, perspectives presented by Anthis et al. \cite{anthisposition} and Wu et al. \cite{wu2025llm} offer caution. The latter emphasises the importance of evaluating simulations based on their ability to reproduce collective patterns rather than individual behaviours. This paper advances this view by highlighting the differences in collective action strategies in various simulation approaches. LLM-based simulation has also been explored in game-theoretic contexts (see \cite{feng2024survey} for a survey), although most existing studies have focused on canonical games rather than comprehensive real-world applications.

\vspace{-0.5em}
\section{Conclusion}
In this paper, we examined the integration of Large Language Models (LLMs) with Empirical Game-Theoretic Analysis (EGTA) as a means of simulating complex social-ecological systems (SES). Using four distinct modelling frameworks (procedural ABMs, generative ABMs, naive LLM-EGTA, and expert-guided LLM-EGTA) applied to a real-world case study of irrigation and fishing in the Amu Darya basin, we identified two central findings. First, the manner in which LLMs are integrated into the simulation pipeline fundamentally shapes the resulting system dynamics. The outcomes ranged from rapid resource collapse in the naive LLM-EGTA models to long-term community survival in the expert-guided variants. Naive implementations failed to capture the underlying structure of the SES interaction game, often producing abstract or degenerate strategy spaces. Second, an expert-guided approach proved essential for stability and economic diversity. The expert-guided LLM-EGTA framework was the only decentralized model capable of sustaining a mixed agriculture–fishing economy under moderate Pigouvian taxation ($\tau = 0.25$). Although prompt-based behavioural induction (e.g., rational, balanced, or altruistic profiles) affected model behaviour, it was insufficient to achieve sustainable equilibria without explicit mechanisms that internalize social costs. Overall, these findings suggest that the formal structure and control afforded by a human expert-in-the-loop methodologies are essential in a LLM-augmented simulation for robust institutional analysis in social-ecological systems.




\bibliographystyle{ACM-Reference-Format} 
\bibliography{sample}


\end{document}